\documentclass[letterpaper, 10pt, conference]{ieeeconf}    
\IEEEoverridecommandlockouts                 
\usepackage{microtype}
\usepackage{verbatim}
\usepackage[singlespacing]{setspace}
\usepackage{hyperref}
\usepackage{balance}
\usepackage{amssymb}
\usepackage{multirow}
\newcommand{\DMfull}{decision maker\xspace}
\newcommand{\DM}{DM\xspace}

\newcommand{\SP}{secretary problem\xspace}

\newcommand{\SSPfull}{Sequential Selection Problem\xspace}
\newcommand{\SSP}{SSP\xspace}
\newcommand{\SSPs}{SSPs\xspace}

\newcommand{\WSSPfull}{Warm-starting \SSPfull}
\newcommand{\WSSP}{WSSP\xspace}

\newcommand{\MSSPfull}		 {Multi-round Sequential Selection Problem\xspace}

\newcommand{\sample}    		 {\sample\xspace}

\renewcommand{\sample}    	 {sample\xspace}

\newcommand{\CCMfull}    	 {Cutoff-based Cost Minimization\xspace}
\newcommand{\CCM}    	 		 {CCM\xspace}

\newcommand{\CCMDPfull}			   {Warm-starting Dynamic Thresholding\xspace}
\newcommand{\CCMDP}			   {WDT\xspace}
\newcommand{\MEAN}			   {MEAN\xspace}

\newcommand{\cost}    		 {\phi}

\newcommand{\nres}		{r}

\newcommand{\captionSize}    {\footnotesize}

\newcommand{\score}			{score\xspace} 
\newcommand{\scores}			{scores\xspace}

\newcommand{\val}			      {S}
\newcommand{\Sbold}	               {\mathbf{\val}} 

\renewcommand{\S}[1]				{ \val_{#1} }

\newcommand{\symbolpres}		   {dot\xspace}
\newcommand{\symb}		{\dot}
\newcommand{\Spres}			{\symb{\val}} 

\newcommand{\Apres}			{\symb{A}}
\newcommand{\Sobold}		  	{ \symb{\mathbf{\val}} }
\newcommand{\So}[1]			 {\Spres_{(#1)}}

\newcommand{\thresh}[1] 			{T_{#1}}

\newcommand{\Abold}		        { \mathbf{A} } 
\newcommand{\A}[1]				{A_{#1}}
\newcommand{\Aobold}[1]	              {\symb{\mathbf{A}}_{#1}} 
\newcommand{\Ao}[2]		       {\Apres_{(#1),#2}}

\newcommand{\Xj}	{X}
\newcommand{\Yj}	{Y}

\newcommand{\regret}   {regret\xspace}
\usepackage{graphics}
\usepackage{trimclip}
\usepackage{algorithm}
\usepackage{algorithmicx,algpseudocode}

\newtheorem{proposition}{Proposition}
\newtheorem{definition}{Definition}
\newtheorem{remark}{Remark}

\usepackage{xr}
\usepackage{hyperref}
\usepackage{url}
\usepackage{amsmath}
\usepackage{amssymb}
\usepackage{dsfont}
\usepackage{upgreek}
\usepackage{bbm}		
\usepackage{graphicx}
\usepackage{epstopdf}
\graphicspath{ {./figures/} } 
\usepackage{tikz}
\usepackage{float}
\usepackage{booktabs} 
\usepackage{subfigure}
\usepackage{xspace}
\usepackage[scaled=.7]{beramono}

\usepackage{tikz}
\DeclareRobustCommand\sampleline[1]{%
  \tikz\draw[#1] (0,0) (0,\the\dimexpr\fontdimen22\textfont2\relax)
  -- (1.4em,\the\dimexpr\fontdimen22\textfont2\relax);%
}

\makeatletter
\g@addto@macro\bfseries{\boldmath}
\makeatother
\newcounter{phase}[algorithm]
\newlength{\phaserulewidth}
\newcommand{\setphaserulewidth}{\setlength{\phaserulewidth}}

\makeatother
\setphaserulewidth{.3pt}

\newcommand{\Sec}[1]		{Section\,\ref{#1}}
\newcommand{\Fig}[1]		{Fig.\,\ref{#1}}

\newcommand{\Eq}[1]			{Eq.\,\ref{#1}}
\newcommand{\Tab}[1]		{Tab.\,\ref{#1}}
\newcommand{\Alg}[1]		{Algorithm\,\ref{#1}}

\newcommand{\Proposition}[1]{Proposition~\ref{#1}}

\newcommand{\Definition}[1]{Definition~\ref{#1}}

\newcommand{\Remark}[1]{Remark~\ref{#1}}

\newcommand{\ie}   			{i.e.\@\xspace}
\newcommand{\eg}   			{e.g.\@\xspace}
\newcommand{\etc}   		{etc.\xspace}

\newcommand{\iid}   		{i.i.d.\@\xspace}

\newcommand{\Ind}[1]    {\mathds{1}{\{#1\}}}

\newcommand{\Exp}[1]    {\mathbb{E}[#1]}
\newcommand{\real}      {\mathbb{R}}
\newcommand{\nat}       {\mathbb{N}}

\newcommand{\mydots} 	{...}
\newcommand{\algComment}[1] 	{\hfill/\!/\,{#1}}

\newcommand{\inlinetitle}[2]  {\vspace{4pt}\noindent\textbf{\emph{#1}{#2}}}

\newcommand{\obullet}				 {\square}
\newcounter{marginNoteCounter}

\begin{document}
\title{\textbf{Optimal Multiple Stopping Rule for Warm-Starting Sequential Selection}}
\date{}
\author{Mathilde Fekom \quad Nicolas Vayatis \quad  Argyris Kalogeratos
\thanks{\!\!\!\!\!\!$\obullet$~The authors are with Universit\'{e} Paris-Saclay, ENS Paris-Saclay, CNRS, Centre Borelli, F-94235, Cachan, France. Emails:~
{\tt\footnotesize  \{fekom,\,vayatis,\,kalogeratos\}@cmla.ens-cachan.fr}.}
\thanks{\!\!\!\!\!\!$\obullet$~Part of this work was funded by the French Railway Company, SNCF, and the IdAML Chair hosted at ENS Paris-Saclay.
}
}

\maketitle
\begin{abstract}
In this paper we present the \emph{\CCMDPfull} algorithm, developed using dynamic programming, for a variant of the standard online selection problem. The problem allows job positions to be either free or already occupied at the beginning of the process. Throughout the selection process, the decision maker interviews one after the other the new candidates and reveals a quality score for each of them. Based on that information, she can (re)assign each job at most once by taking immediate and irrevocable decisions. We relax the hard requirement of the class of dynamic programming algorithms to perfectly know the distribution from which the scores of candidates are drawn, by presenting extensions for the partial and no-information cases, in which the decision maker can learn the underlying score distribution sequentially while interviewing candidates.
\end{abstract}

\section{Introduction}
Sequential Selection Problems (\SSPs) occur in several real-life situations. A common characteristic of such decision processes is that decisions need to be \emph{immediate} and \emph{irrevocable}. 
For instance, at an emergency healthcare unit, patients arrive sequentially seeking for medical help \cite{Gnanlet09}. Other examples are found in kidney exchanges \cite{Chisca18}, auctions \cite{Kleinberg05}, robotic sampling \cite{robotic2015}, and most straightforwardly in recruitment processes that involve sequential interviews of job-seekers by a decision maker (DM), who immediately takes hiring or rejection decisions \cite{Broder09}. For clarity, we adopt the terminology of the latter problem, \eg \emph{candidates} are interviewed, \emph{hiring} means selecting, \emph{firing} is deselecting, \etc

In the most well-known \SSP, the \emph{\SP} \cite{Ferguson89, Lindley61}, $n \in \nat^*$ candidates are sequentially interviewed for a single job position, and a hire puts an end to the recruitment process. Its multi-choice extension, also called \emph{multiple stopping problem}, has been extensively studied \cite{Kleinberg05, Bearden06bis, Babaioff07}, and terminates when the desired number of $b\in \nat^*$ hired have been decided. In the \emph{hiring problem} \cite{Broder09}, the \DM's task is to grow the company as much as possible while keeping maximal the average \score of the hired employees; therefore, there is no limit in the number of job positions. %
The \emph{hiring-above-the-mean} policy was proposed, which selects a candidate when his score exceeds the average score of the current employees (changes over time). This policy is an efficient and easy to implement, yet notably dependent on the score distribution and sensitive to the quality of the first hires.

What we call as the \emph{full-information} case, assumes that the distribution from which the scores of candidates are sampled is known to the \DM. In \cite{Nikolaev07}, dynamic programming  is employed to compute the optimal stopping rule for each candidate, \ie the score threshold to beat in order to be hired. This threshold depends on his \emph{arrival time} and on the number of jobs positions that remain to be filled.

One important limitation of all the aforementioned settings is that they begin with an empty selection set. The case of performing a selection starting with a set of items at hand was only recently introduced in \cite{Fekom19}. There, a \emph{preselection} is composed of the pre-existing items that can then be \emph{replaced}, by selecting better items from those that appear sequentially, to produce the final selection. 
More specifically, the authors present the so-called \emph{\WSSPfull} (\WSSP) where the \DM can perform at most one update per job position. This setting can also suit well to multi-round applications where the output of a round is the input of the subsequent one.

The \emph{\CCMfull} algorithm from the same work \cite{Fekom19} makes no assumption on the distribution of candidates' scores. It splits the process into two phases: the \emph{learning phase} in which the \DM learns an acceptance threshold from candidates that are getting rejected by default, and thereafter, the \emph{selection phase} which uses that threshold to accept or reject from the rest candidates. Despite its biased rule, which is due to the disregard of the early-arriving candidates, the strategy can be efficient and robust to score changes, provided an appropriately set learning phase size. 

Important to note, when the score distribution is known, the cutoff-based strategies are not optimal and policies derived from dynamic programming are more suitable. To the best of our knowledge, there exists no such warm-starting strategy in the literature.

\inlinetitle{Contribution}{.} Inspired by the warm-starting aspect of the \WSSP \cite{Fekom19}, we propose the \emph{\CCMDPfull} (\CCMDP) algorithm that attributes to each incoming candidate a threshold value to beat according to: i) its arrival time, ii) the current number of empty job positions, and iii) the current number of positions occupied by initial employees (\ie available employees that can be replaced).  
The threshold value to beat is computed by means of dynamic programming, adapted to the warm-starting scenario. The algorithm is easy to implement and gives each candidate a chance to be hired regardless his arriving time. \CCMDP's downside lies in the assumption that the score distribution is known. We relax this requirement by first considering the \emph{partial information} case where the \DM only knows the nature of the distribution, and then the \emph{no-information} case where the threshold is purely rank-based. We show with simulations that the \CCMDP outperforms existing methods in the warm-starting scenario.

\newpage
\section{Setting}\label{sec:setting}

The \SSPfull (\SSP) of interest is the following: $n \in \nat^*$ candidates are sequentially interviewed by a \DMfull (\DM) who has the task of managing a limited budget of $b\in \nat^*$, $b\leq n$, interchangeable job positions. Initially, $\nres\leq b$ positions are empty as some employees have become permanently unavailable, while the rest $b-\nres$ are initially occupied by existing employees that form a set called \emph{preselection}. Available and unavailable employees constitute a \emph{reference set} for the \DM. Once candidate $j$ arrives, the \DM observes his score $S_j \in \real$, which, like all scores, it is assumed to follow a known distribution $f_S$. In our notations, an added \symbolpres on top of a variable refers to the preselection, \eg $\Sobold = (\So{1},\So{2},...,\So{b-r})$ gives the score of each preselected employee. For convenience, the latter vector is considered to be sorted in decreasing order, \ie $\So{1}$ and $\So{b-r}$ are respectively the best and the worst scores of the preselected employees.
Here are the basic rules of the game:
\begin{itemize}
\item[R1.] Each decision (\ie hire or not) shall be immediate and irrevocable, which means that every position can be assigned (or reassigned) at most once throughout the selection process.
\item[R2.] The \DM must at least fill the empty job positions by hiring $\nres$ candidates.
\item[R3.] Once no empty position is left, the \DM can still hire an incoming candidate  by removing one of the employees of the preselection from his position (only fire on hire).
\end{itemize}

The following definition is a score-based adaptation of the \WSSPfull (\WSSP) proposed in \cite{Fekom19}, where here the \DM initially knows also the distribution from which candidates' scores are drawn. 
\vspace{2mm}
\begin{definition}{\emph{Distribution-aware \WSSP}} is the online selection process described by elements of three categories:
\\
\noindent \textit{1)~Background}

$\mathcal{B} =(b,n,f_S,\Sobold)$: collection of elements initially at \DM's knowledge or disposition, including:

\begin{itemize}
\item $n \in \mathbbm{N}^*$: finite number of candidates to appear;
\item $b \in \mathbbm{N}^*$, $b \leq n$: number of resources;
\item $f_S$: the distribution generating candidates' scores;
\item $\Sobold = (\So{1},\mydots,\So{b-r}) \in \real^{b-r}$; $\So{j} \sim f_S,\ \forall j$: \scores of the preselection.
\end{itemize}
2)~\textit{Process} 
\begin{itemize} 
\item $\Sbold = (\S{1},\mydots,\S{n}) \in \real^{n}$; $\S{j} \sim f_S,\ \forall j$: candidates' scores;
\item $\Abold = (\A{1},\mydots,\A{n}) \in\{0,1\}^{n}$: sequence of decisions for candidates, \ie $\A{j} =1$ if the $j$-th candidate gets hired. 
\end{itemize}
3)~\textit{Evaluation }
\begin{itemize}
\item The \emph{reward} is evaluated at the end of the process by:
\begin{equation}
 \phi_{\mathcal{B}} (\Sbold,\Abold)
 = \sum_{i=1}^{b-r} \So{i} \Ao{i}{n} + \sum_{j=1}^{n}  \S{j} \A{j}\, \, \ge 0,
 \end{equation}
where $\Ao{i}{n}\in\{0,\,1\}$ indicating if the $i$-th preselected employee kept his position after $n$ candidate interviews.
\end{itemize}
\label{def:wssp}
\end{definition}

The evaluation criterion to maximize is therefore the expectation of the above reward function, $\Exp{\phi_{\mathcal{B}} (\Sbold,\Abold )}$. 
Note that, since the \DM is forced to fill the $b$ positions by the end of the process, it holds: $\sum_{i=1}^{b-r}\Ao{j}{n} + \sum_{j=1}^{n}\A{j} = b$.

\section{\CCMDPfull}
Here, we present the \emph{\CCMDPfull} (\CCMDP) method to solve optimally the distribution-aware \WSSP of \Definition{def:wssp}. 
Without loss of generality, we consider non-negative \iid scores $S_j \geq 0, \, \forall j$, and that the best- (resp. worst-) skilled individual has the highest (resp. lowest) score. 

\subsection{Threshold-based algorithm}
The idea behind the strategy is to find the optimal \emph{acceptance threshold} $\thresh{j} \in \real$ that the $j$-th arriving candidate should beat in order to be hired, $\forall j \in \{1,...,n\}$; see the example of \Fig{fig:dynamic_prog}. Note that, in order to be optimal, this threshold must depend on the state of the ongoing selection process. Therefore, we write $T_j = T_j^{X_j,Y_j}$, where $X_j \in  \{0,...,r\}$ positions are still empty and $Y_j \in \{0,...,b-r\}$ jobs are still occupied by preselected employees, \ie after $j-1 \in \nat^*$ interviews and while the $j$-th candidate is being interviewed. 
Using the notations of \Definition{def:wssp} we get: 
\begin{equation}
X_j := \max\left(r- \sum_{i=1}^{j-1} A_i,\,\,0\right)\, \ \text{and}\, \ Y_j := \sum_{i=1}^r \Ao{i}{j}.
\end{equation}
To simplify our notations, we omit the dependency of $X_j$ and $Y_j$ to $j$, and we simply write $X$ and $Y$ to refer to their value at the implied step $j$ of the selection process.

\begin{figure}[t]
\centering
{\includegraphics[width = 0.94\linewidth,viewport= 60 110 870 700, clip]{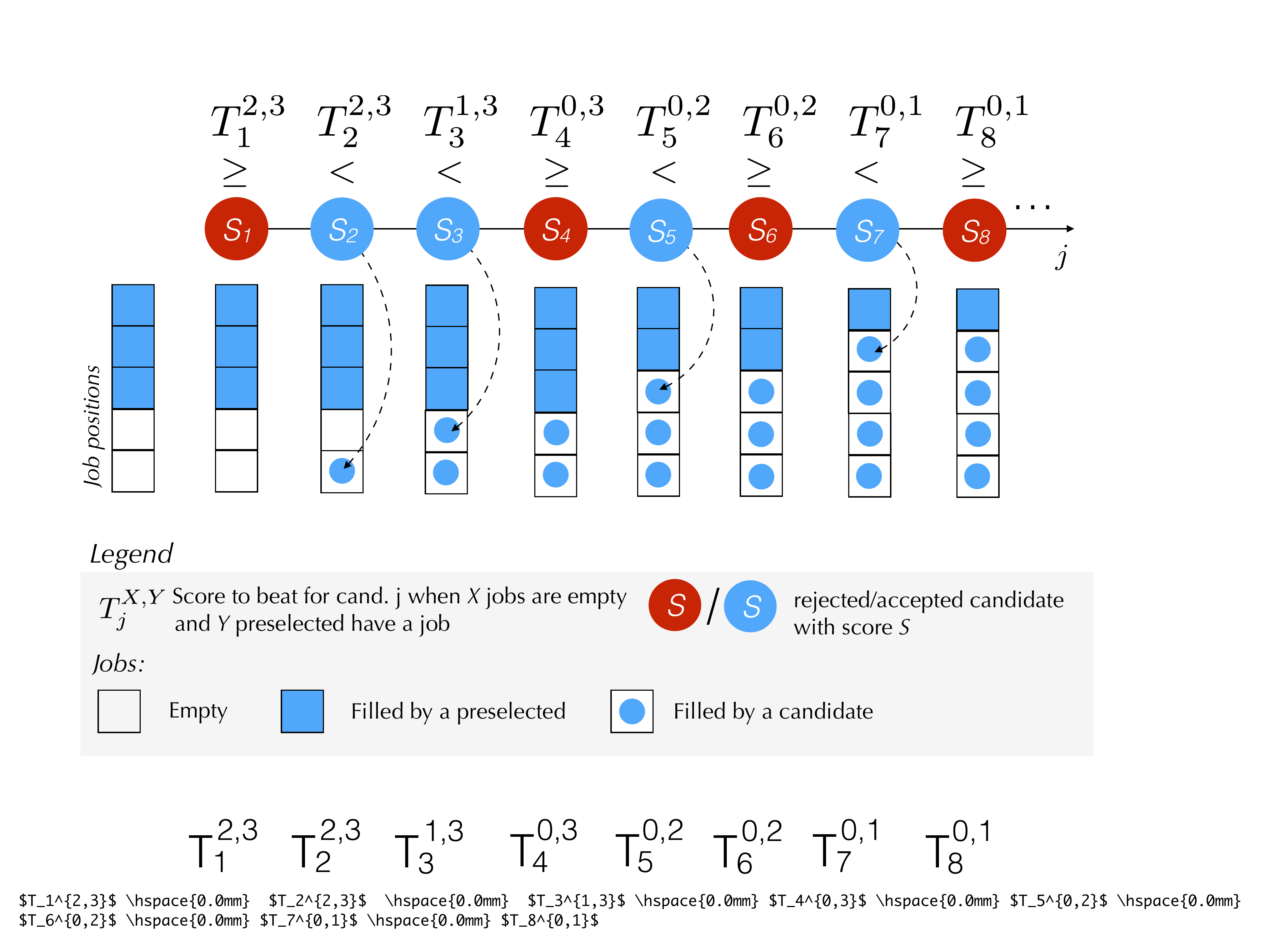}} 
\vspace{-5mm}
\caption{Demonstration of the \CCMDP algorithm. The score of an incoming candidate, $S_j$, is compared to an associated acceptance threshold $T_j^{\Xj,\Yj}$\!, where $\Xj$ and $\Yj$ are respectively the current number of empty positions and that of positions occupied by preselected employees. Accepted candidates (blue circle) first fill empty positions and then, if they are competitive enough, take a preselected employee's positions (\eg the $5$-th candidate).}
\label{fi:dyn_prog}
\end{figure}

\inlinetitle{Value function}{.} The fundamental question remains how to compute the thresholds optimally. The solution can be found via dynamic programming where the \emph{value function}, which is the expected value of the \regret here, is computed for each possible scenario. Let $V_j^{\Xj,\Yj}\!\in \real$ be this value function when $j-1 \in \nat^*$ candidates have been interviewed so far. 

Before going into the method's details, we can deduce the following remarks from the rules that constrain the process:
\begin{remark} The second rule, R2, implies that
if $\Xj\geq n-j$, then $V^{\Xj,\Yj}_j = 0, \, \forall \, \Yj$. Thus, the last incoming candidates might be accepted by default, and this leads to $\Xj = 0$ when $j=n$. 
\label{rem:empty}
\end{remark}
\begin{remark} R3 implies that if $\Xj \neq 0$, then $\Yj = b-r, \, \forall j$.
\end{remark}
\begin{remark} From R2 and R3 we deduce that $\Xj$ and $\Yj$ are non-increasing with $j$: throughout the process the number of job positions assigned to candidates cannot decrease, while those for the preselected employees cannot increase.
\end{remark}

To compute $V_j^{\Xj,\Yj}$ we work by means of backward induction. First, we consider the extreme case that $n$ candidates are automatically rejected, and compute the value function $V_n^{r, b-r}$. Here, either $X=0$ and $r=0$,\, \ie each position is occupied by a preselected employee, or $V_n^{r, b-r}=0$ because one or more positions are empty. The induction then goes to the first non-zero value function, which is when $n-r$ candidates get automatically rejected: there, the only option is to hire the $r$ last candidates to arrive (see \Remark{rem:empty}), hence:
\begin{equation} 
V_{n-r+1}^{r,b-r} = \mathbbm{E}\left[{\sum_{j=n-\nres+1}^{n} \!\!S_j + \sum_{i=1}^{b-\nres}\So{i}}\right] = \nres \mu +  \sum_{i=1}^{b-\nres}\So{i},
\label{eq:first_value}
\end{equation}
where $\mu$ is the mean of the known score distribution $f_S$.

One step back, $V_{n-r}^{r,b-r}$ is evaluated by accounting every option: either the $(n-r)$-th candidate gets rejected and the last $r$ candidates get hired, leading to the reward of $V_{n-r+1}^{r,b-r}$ (\Eq{eq:first_value}), or the $(n-r)$-th candidate gets hired and in this case we should, again, present two options. The \DM can lower by one either the number of empty job positions, or the number of preselected employees that will keep their positions at the end of the selection. This reasoning is generalized to the following recurrent value functions $\forall j \in \{1,...n\}$:
\vspace{2mm}
\begin{align}
\begin{split}
\!\!\!\!\!\!\!\!\!V^{\Xj,\Yj}_{j}\! &= \mathbbm{E}\left[\max\left(V^{\Xj,\Yj}_{j+1}\!\!,\ S_{j}+ \max(V^{\Xj-1,\Yj}_{j+1}\!\!, V^{\Xj,\Yj-1}_{j+1})\right)\right]\!\!.\!\!\!\!\!\!\!\!
\end{split}
\label{eq:recurrence}
\end{align}
The first term in the outer $\max(\cdot)$ corresponds to the option of rejecting the $j$-th candidate, while the second term is the option of accepting him.
Two observations can be made: i)~if the goal is to minimize instead of maximize the objective, then $\max(\cdot)$ should be replaced by $\min(\cdot)$ functions; 
ii)~results of \cite{Nikolaev07}, and specifically their Theorem~3 in Sec.~3, can be retrieved  by setting $Y=0$ that implies $V_{j+1}^{X,Y-1}=0$.

\inlinetitle{Recurrence relation}{.}
Thanks to our backward induction formulation, we enunciate a generic formula for the recurrence relation of the value function.

\begin{proposition} Let a \WSSP with a population of \iid scores $S \in [\alpha, \beta]$, each drawn from a known distribution $f_S$ with cumulative distribution function $F_S(x) = \int^x_\alpha f_S(y)dy$. Having processed $j-1$ interviews, $\Xj \in \{0,...,r\}$ job positions are empty and $\Yj \in \{0,...,b-r\}$ are occupied by preselected employees.
Then, the recurrence relation of \Eq{eq:recurrence} becomes: 
\begin{equation}
\!V^{\Xj,\Yj}_{j}\!\!- V^{\Xj,\Yj}_{j+1} = Z_{j+1}^{\Xj,\Yj}\!\big(F_S(Z_{j+1}^{\Xj,\Yj})-1\big) +\!\!\int_{Z_{j+1}^{\Xj,\Yj}}^{\beta}\!s f_S(s) ds,\!\!\!\!\!\!\!\!\!\!\!\!
\end{equation}
where $Z_{j+1}^{\Xj,\Yj} := V^{\Xj,\Yj}_{j+1} - \max(V^{\Xj-1,\Yj}_{j+1}, V^{\Xj,\Yj-1}_{j+1})$.
\label{prop:relation}
\end{proposition}
The proof\footnote{A separate Appendix, containing detailed proofs, can be found as a supplement file.} uses ${\max(A,B)} = {A\Ind{A>B} +B\Ind{A \leq B}}$ in order to compute the expectation.

\vspace{2mm}
\inlinetitle{Optimal threshold}{.} Evidently, the $j$-th candidate must be accepted if 
the value function associated to the option of accepting him is larger than that of rejecting him. According to \Eq{eq:recurrence}, the latter amounts to choosing the optimal threshold defined in the following proposition that maximize the expectation of the reward. 
\begin{proposition} Let a \WSSP where $j-1$ interviews have been processed, $\Xj \in \{0,...,r\}$ job positions are empty, and $\Yj \in \{0,...,b-r\}$ are occupied by preselected employees. The optimal acceptance threshold for candidate $j$ is defined as:
\begin{equation}
T_{j}^{\Xj,\Yj} =V^{\Xj,\Yj}_{j+1} - \max(V^{\Xj-1,\Yj}_{j+1}, V^{\Xj,\Yj-1}_{j+1}).
\vspace{1mm}
\end{equation}
Essentially, the $j$-th candidate is accepted if his score beats the corresponding threshold, \ie $A_j = \Ind{S_j> T^{\Xj,\Yj}_j}$.
\label{prop:threshold}
\end{proposition} 

The \CCMDP procedure works with the optimal threshold described above and is described in \Alg{alg:ccmdp}. 

\begin{algorithm}[t]
\footnotesize
\caption{\CCMDPfull (\CCMDP)}
{\bf Input:} 
the evaluation table of the value function $V_j^{\Xj,\Yj}$ where $b$, $r$, and $n$ are the numbers of resp. job positions, initially empty among them, and sequentially incoming candidates; $\Aobold{0} =(\Ao{1}{0},...,\Ao{b-r}{0}) = (1,...,1)$ is the initial status of the preselected employees. 

{\bf Output:} the set of final job assignments $\Aobold{n}  \in \{0,1\}^{b-r}$, and $\Abold \in \{0,1\}^{n}$.
\begin{algorithmic}[1]
\State $X\leftarrow r$\ \algComment{nb. of empty jobs}
\State $Y \leftarrow b-r$ \algComment{nb. of jobs occupied by preselected employees}
\For{$j = 1$ to $n$}
\State $T_j \leftarrow V^{\Xj,\Yj}_{j+1} - \max(V^{\Xj-1,\Yj}_{j+1}, V^{\Xj,\Yj-1}_{j+1})$ \algComment{see \Proposition{prop:threshold}}
\If {$\S{j} > T_j$}
\State $A_j \leftarrow 1$\algComment{accept candidate}
\If { $X > 0$ }
\State {$X \leftarrow \max(X-1,0)$}\algComment{fill one empty position}
\Else 
\State ${\Ao{Y}{j} \leftarrow 0}$\algComment{remove job from a preselected employee}
\State $Y \leftarrow Y-1$
\EndIf
\Else
\State $\A{j} \leftarrow 0$
\EndIf
\EndFor
\end{algorithmic}
\label{alg:ccmdp}
\end{algorithm}

\begin{table*}[t]
\scriptsize
\centering
\begin{tabular}{ll|llllllllllllll}
\toprule
$\hspace{-1mm}Y$ &$\hspace{-1.7mm}X$ & $\hspace{-1mm}V_1^{X,Y}$ & $V_2^{X,Y}$ & $V_3^{X,Y}$ & $V_4^{X,Y}$ &$V_5^{X,Y}$ & $V_6^{X,Y}$ & $V_7^{X,Y}$ & $V_8^{X,Y}$& $V_9^{X,Y}$& $V_{10}^{X,Y}$& $V_{11}^{X,Y}$& $V_{12}^{X,Y}$& $V_{13}^{X,Y}$& $V_{14}^{X,Y}$\\ 
\midrule
\midrule
\multirow{2}{*}{\begin{tabular}[c]{@{}l@{}} \hspace{-1mm}0 \end{tabular}}&\hspace{-1.5mm}1 & \hspace{-1mm}0.893 & 0.886 & 0.879 & 0.871 & 0.861 & 0.850 &0.836 &0.823 & 0.800 &0.775 &0.741&0.768 & 0.732 & 0.682\\
& \hspace{-1.5mm}2 & \hspace{-1mm}1.719 & 1.702 & 1.683 & 1.661 & 1.636 & 1.606 &1.571 & 1.529 & 1.476 & 1.409 &1.320 & 1.195 & 1.000 & 0.000\\
\midrule
\multirow{3}{*}{\begin{tabular}[c]{@{}l@{}} \hspace{-1mm}1 \end{tabular}}&\hspace{-1.5mm}0 & \hspace{-1mm}0.907 & 0.902 & 0.897 & 0.891 & 0.885 & 0.877 & 0.869 & 0.859 & 0.847 &  0.833 & 0.816 &0.795 & 0.979 & 0.979 \\
& \hspace{-1.5mm}1& \hspace{-1mm}1.756 & 1.742 & 1.729 & 1.712 & 1.694 & 1.673 & 1.650 & 1.621 & 1.588 & 1.547 & 1.495 & 1.428 & 1.333 & 1.182\\ 
& \hspace{-1.5mm}2& \hspace{-1mm}2.547 & 2.523 & 2.496 & 2.465 & 2.431 & 2.391 & 2.345 & 2.290 &  2.224 & 2.142 & 2.036 & 1.894 & 1.682 & 0.000\\
 \bottomrule
\end{tabular}
\normalsize
\caption{\captionSize Evaluation of the $V_j^{\Xj, \Yj}$ function at each step $j=1,...,14$. The distribution of the scores is $S_j \sim \mathcal{U}(0,1)$, and there is $b-r=2-1=1$ initially non-empty job position occupied by an employee with score $\So{1}=0.682$.}
\vspace{-6mm}
\label{tab:value_function}
\end{table*}

\subsection{Special case: uniform distribution}\label{sec:unif}
Consider that scores are uniformly distributed in $[\alpha, \beta]$. The cumulative and probability density functions being easily computed, allows the recurrence relation to get simplified.
\begin{proposition} Set $S_j \sim \mathcal{U}(\alpha,\beta), \, \forall j \leq n$. Then, \Proposition{prop:relation} becomes:
\begin{equation}
V^{\Xj,\Yj}_{j} - V^{\Xj,\Yj}_{j+1} = \frac{{Z^{\Xj,\Yj}_{j+1}}^2- 2\alpha Z^{\Xj,\Yj}_{j+1}+\beta^2 }{2(\beta-\alpha)} - Z^{\Xj,\Yj}_{j+1},
\end{equation}
where $Z_{j+1}^{\Xj,\Yj}= V^{\Xj,\Yj}_{j+1} - \max(V^{\Xj-1,\Yj}_{j+1}, V^{\Xj,\Yj-1}_{j+1})$.
\label{prop:unif}
\end{proposition}
\vspace{2mm}
The proof$^{\text{\ref{fnote}}}$ uses the fact that $F_S(x) = \frac{x-\alpha}{\beta-\alpha}$ and $f_S(x) = \frac{1}{\beta-\alpha}$ for a uniform distribution.

\inlinetitle{Example}{.}
Set $\alpha = 0$ and $\beta = 1$, $b=3$, $r=2$, and $n=14$. \Tab{tab:value_function} displays the evaluation of the value function $V_j^{X,Y}$ at each step of the selection process. Recall that, at the respective step $j$, $X \in \{1,...,b\}$ is the number of empty job positions, and $Y \in \{0,\,1\}$ is the \DM's hiring decision. Now, imagine the following sequence of scores: $\Sbold=( 0.498  , 0.858  ,  0.749 ,   0.398 ,  ...)$ and $\Sobold = (0.682)$ is the score of the preselected employee. We are determining the acceptance threshold sequentially. Note that, since $r\neq 0$, $V_j^{0,0} =0$. For the first incoming candidate, $j=1$, $r=2$ job positions are empty, \ie $X=2$ and $b-r=1$ position is occupied by a preselected employee, \ie $Y=1$. The first candidate is rejected since $T_1^{2,1} = V^{2,1}_{2} - \max(V^{1,1}_{2}, V^{2,0}_{2}) = 2.523 - \max(1.742,1.702) = 0.781>0.498$.
The second threshold reads $T_2^{2,1} = V^{2,1}_{3} - \max(V^{1,1}_{3}, V^{2,0}_{3}) = 2.496 -  \max(1.729,1.683) = 0.767<0.858$, which allows the acceptance of the second candidate.
The following threshold is then, $T_3^{1,1} = 0.832$, thus the third candidate is rejected. The process continues this way until the sequence is finished.

\inlinetitle{Rank-based setting}{.} When the score distribution is either unknown (no-information case), or does not exhibit a closed-form cumulative density function, the \DM can only rely on a relative evaluation of the candidates. In practice, 
she can assign a relative rank to each incoming candidate by comparing him to those already examined (let the best one be ranked first). 

In that case, the acceptance threshold is \emph{rank-based}, hence $T_j^{X,Y}$ stands for the \emph{absolute rank} (which cannot be known, though) that the $j$-th candidate needs to exceed to get selected. Conveniently, the absolute ranks of a set follow a discrete uniform distribution that exhibits a closed-form description. Then, the threshold value for the $j$-th candidate is computed using \Proposition{prop:threshold} and, following the same reasoning as in \Proposition{prop:unif}, we get the following simplified expression:

\begin{equation}
V^{\Xj,\Yj}_{j}  = V^{\Xj,\Yj}_{j+1} - \frac{ {Z^{\Xj,\Yj}_{j+1}}^2 -Z^{\Xj,\Yj}_{j+1} }{2(n+b)}, 
\end{equation} where $Z^{\Xj,\Yj}_{j+1} = V^{\Xj,\Yj}_{j+1} - \min(V^{\Xj-1,\Yj}_{j+1}, V^{\Xj,\Yj-1}_{j+1})$.

As mentioned, the \DM cannot know the absolute rank of a candidate before finishing all interviews. She can still, though, estimate it knowing his \emph{relative rank} and by taking into account the proportion of candidates that has already been examined. More precisely, the $j$-th candidate has relative rank denoted by $Z_j^\text{rel} \in \nat^*$ after the examination of $j+b-r$ individuals (including the preselected employees), and the absolute rank denoted by $Z_j^\text{abs}$ that he would have after the examination of $n+b-r$ individuals. 
Hence, we set $Z_j^\text{rel} = \frac{j+b-r}{n+b-r} Z_j^\text{abs}$ and, thereby, the practical threshold 
of relative rank that the $j$-th candidate must exceed to be accepted is:
\begin{equation}
 \tau_j^{\Xj,\Yj} = \frac{j+b-r}{n+b-r}\,T_j^{\Xj,\Yj}.
 \label{eq:rel_rank}
\end{equation}

\section{Simulation results}
\subsection{Simulations parameters}
The \WSSP setting takes as input the reference set (containing available and unavailable employees) and the sequence of candidates, and outputs a selection of size $b$. A very interesting feature of this configuration is that it enables multi-round applications where the output selection of a round can be fed as input for the consecutive round. For instance, it is natural to imagine periodic or reoccurring recruitment processes for large organizations that have several human resources to manage. Inspired by \cite{Fekom19}, we thereby repeat the \WSSP process $K\in \nat^*$ times. Each repetition $k\leq K$, called \emph{round}, is in essence a warm-starting selection. The round starts with $b-r$ available employees and assumes that unavailability occurs uniformly at random among the $b$ employees of the reference set (those selected during the previous round). Overall, the $K$ rounds form a \emph{\MSSPfull} (MSSP) where the \DM's objective is the upkeep of a highly-skilled group of employees throughout the entire process.

In the simulations, a population of $N=10000$ job-seekers is considered, and that $n=100$ interviews take place in each of the $K=10$ considered rounds. The candidates' scores are drawn from a given distribution and remain fixed during the process. The preselected employees of the first round are chosen uniformly at random from the population, and hence, carry an average quality score. We desire to compare our online strategy, the \CCMDP, to the best an offline strategy achieves, \ie in the case that the \DM could examine the candidates altogether as a batch. Therefore, instead of the reward, in the figures we plot the regret defined as $\phi_k = | \phi_{\text{off},k} - \phi_k|, \, \forall k\leq K$, where $\phi_{\text{off},k}$ and $\phi_k$ are respectively the offline and online reward.

\subsection{Score-based setting}

\Fig{fig:dynamic_prog} displays the average \regret $\cost_k$ in different settings, namely $S_{i,k} \sim \mathcal{U}(0,1)$ (top row) and $S_{i,k} \sim \text{Exp}(1), \, \forall i, k$ (bottom row). Let us start by focusing on the plain line curves, one of which is the RAND baseline (grey line) that decides for the hires at random. 
A first straightforward observation is that the average \regret has similar inefficient behavior for both distributions. In the following description, we therefore focus on the uniform distribution.

Secondly, the subfigures on the left assume $r=0$, hence the process always starts with $b$ empty positions, whereas on the right it is assumed $r=b$, thus the process starts with $b$ positions occupied by preselected employees. In the first case, since the employees do not quit their position in-between two subsequent rounds, the \DM cannot deteriorate the selection, and might even improve the set of employees through time by replacing initial employees with more skilled candidates.
The \regret naturally goes to zero, and it does go faster for the proposed \CCMDP than for instance the \MEAN or $\text{\CCM}^{*}$ strategies.  Note that $\text{\CCM}^{*}$ is merely the \CCMfull algorithm with optimal size of learning phase \cite{Fekom19}. In the second case ($r=0$), the \regret does not necessarily converge towards zero; neither \MEAN nor $\text{\CCM}^{*}$ manage to keep a low average \regret. This phenomenon is explained as follows: progressively, the average score of the employees of the previous round gets so competitive for the incoming candidates that it forces the \DM to select the last-arriving ones by default to prevent ending up with having empty positions after all the interviews (see the rule R2 in \Sec{sec:setting}).

\subsection{Full, partial, and no-information settings}

\begin{figure}[t]
\vspace{-7mm}
\hspace{-2mm}
\subfigure{
\clipbox{2pt 0.45pt 3pt 0pt}{\includegraphics[width = 0.51\linewidth,viewport=10 5 570 570, clip]{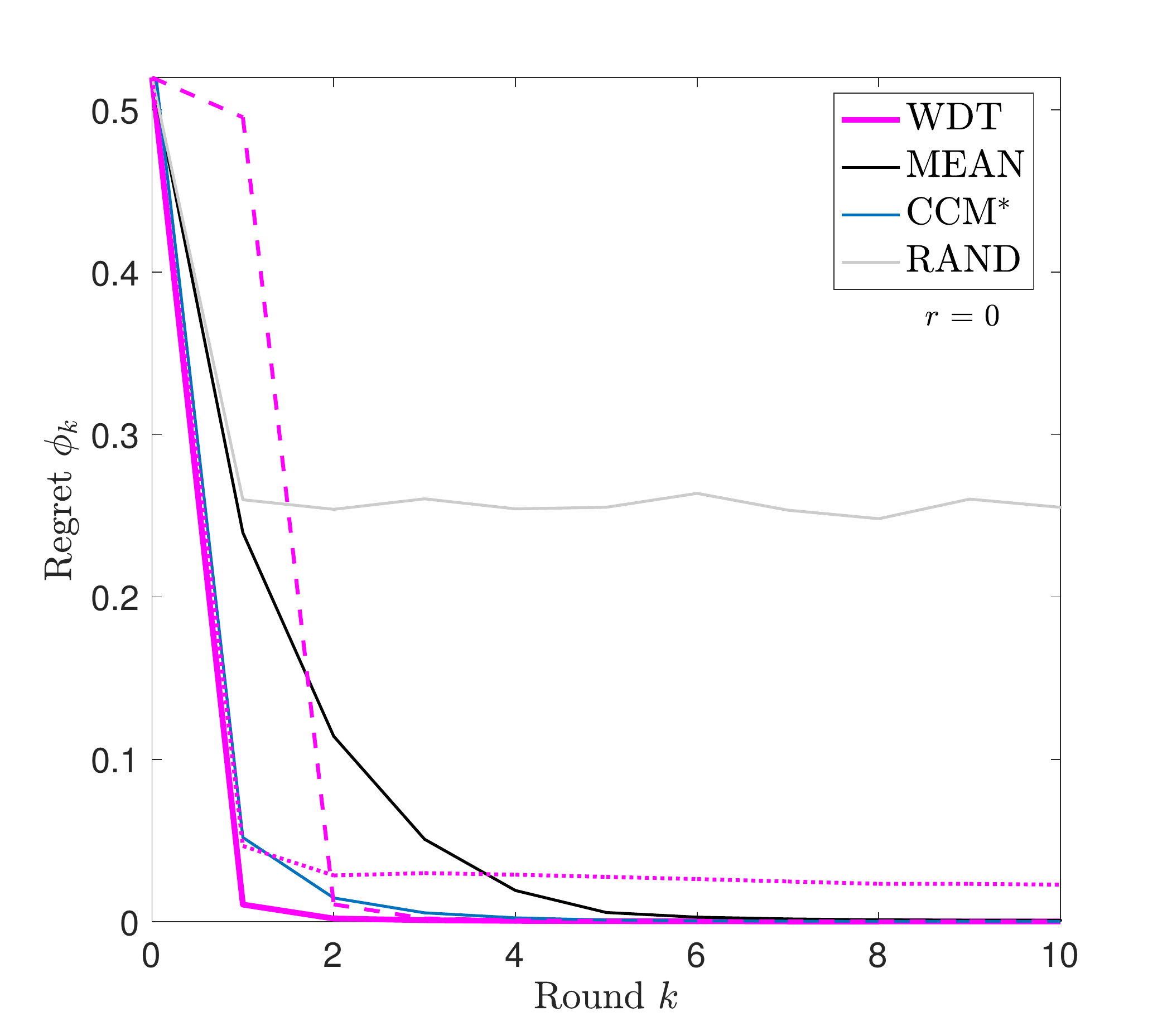}}}
\hspace{-2mm}
\subfigure{
\clipbox{8pt 0.45pt 3pt 0pt}{\includegraphics[width = 0.51\linewidth,viewport=10 5 570 570, clip]{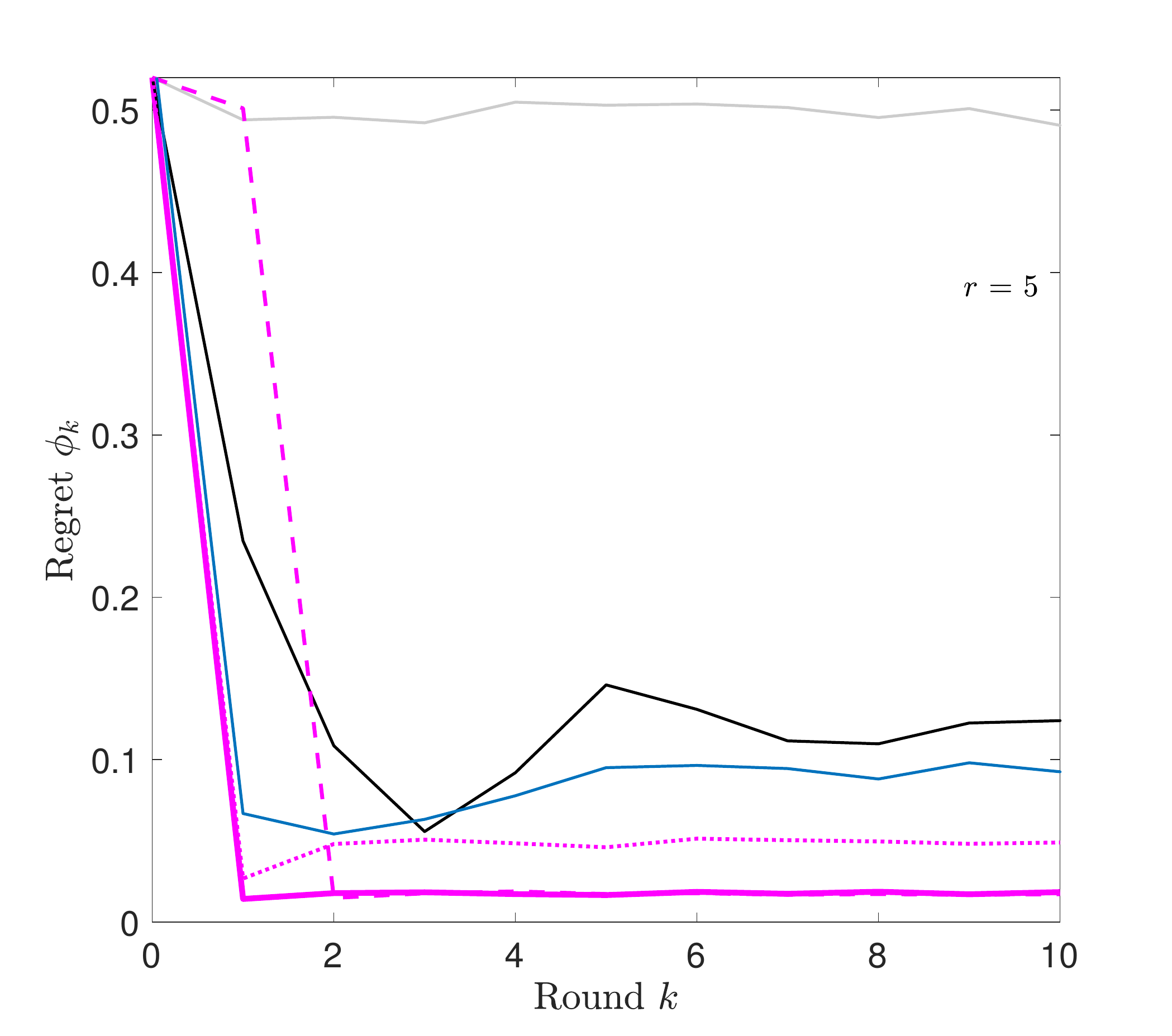}} }\\
\vspace{-9mm}
\\
\subfigure[no resignations ($\nres=0$)]{
\hspace{-2.35mm}
\clipbox{2pt 0.45pt 3pt 0pt}{\includegraphics[width = 0.51\linewidth,viewport=10 5 570 570, clip]{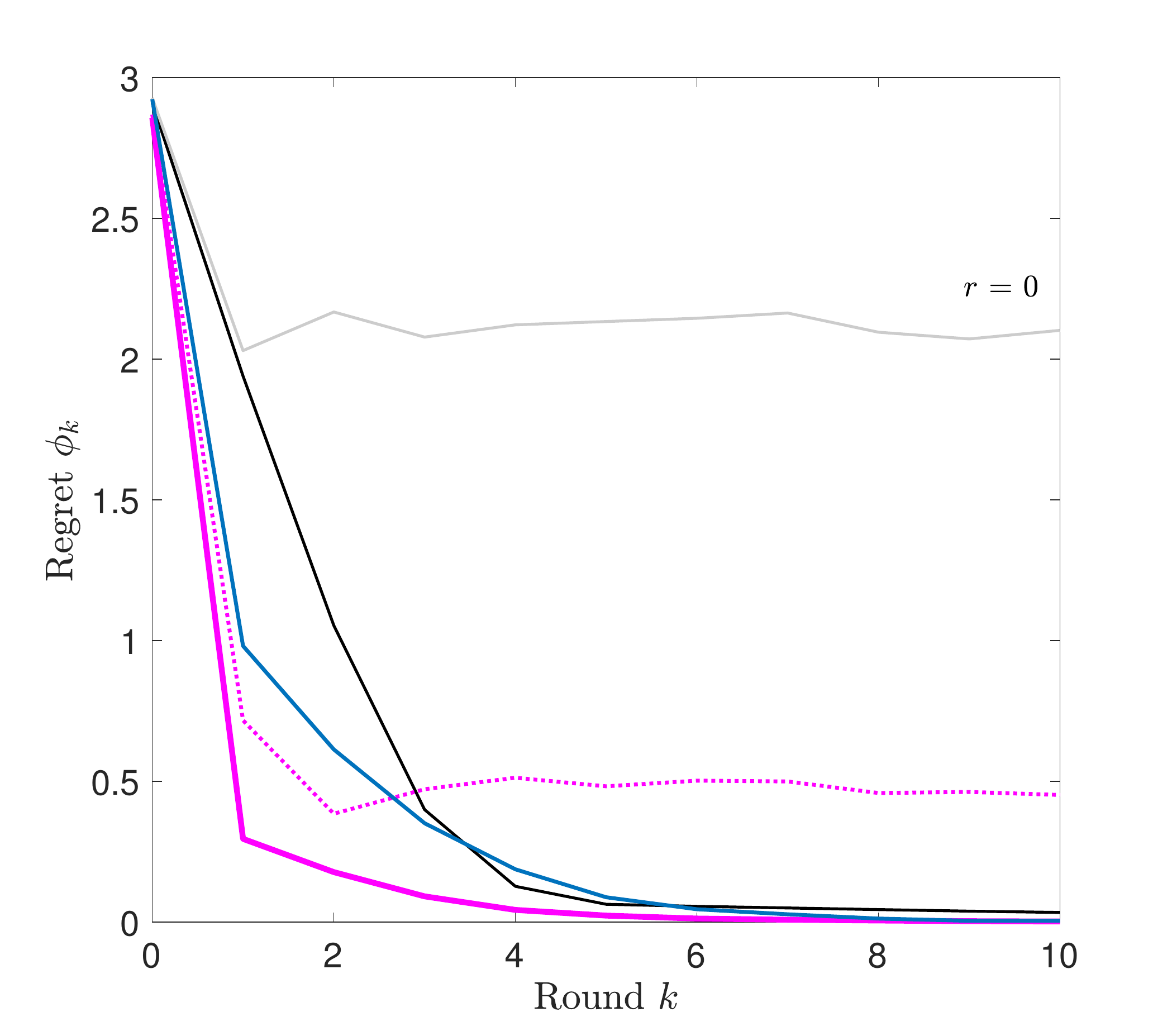} } }
\hspace{-3.75mm}
\subfigure[max resignations ($\nres=b$)]{
\clipbox{8pt 0.45pt 3pt 0pt}{\includegraphics[width = 0.51\linewidth,viewport=10 5 570 570, clip]{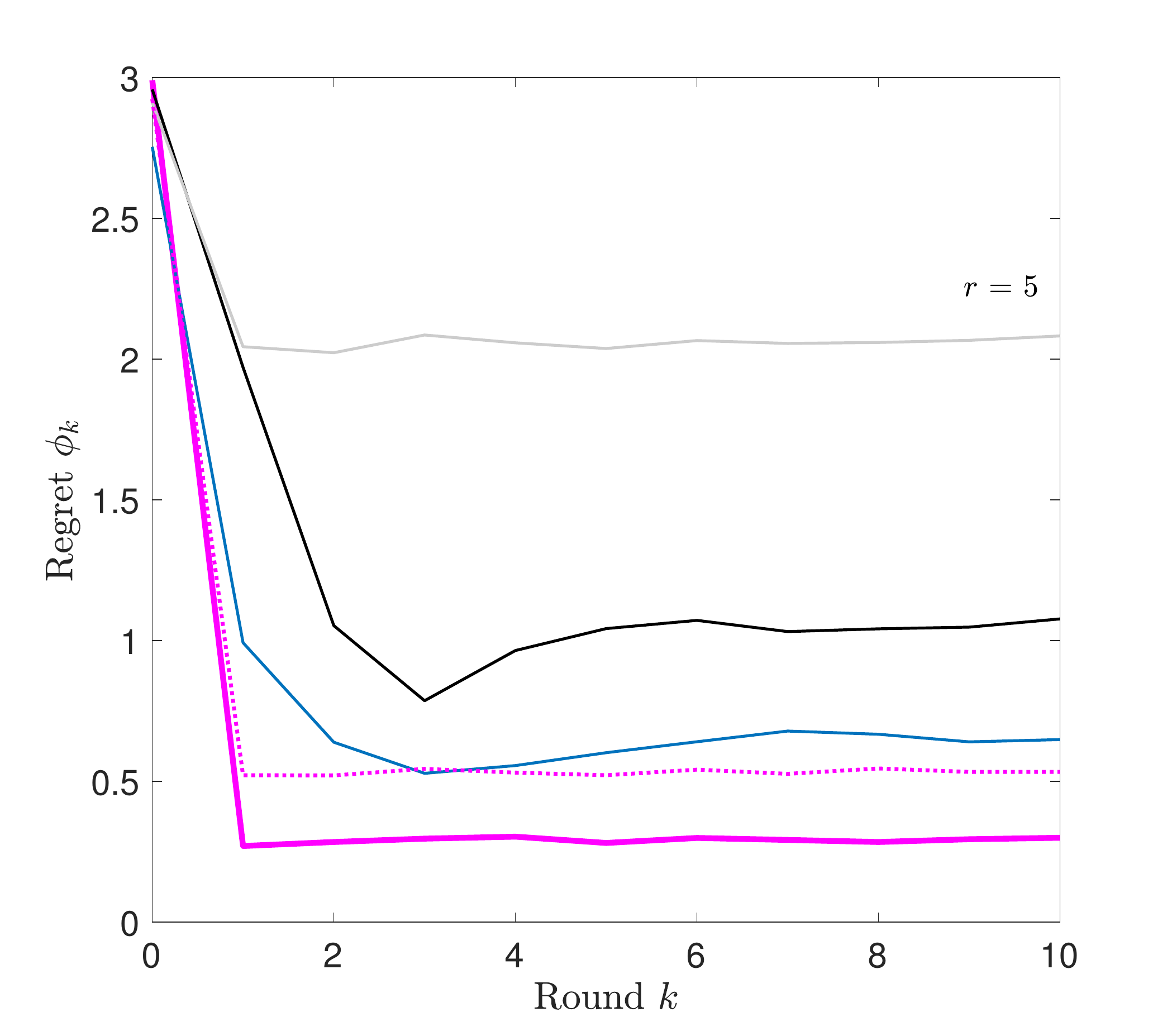}} }
\vspace{-2mm}
\caption{\captionSize Average \regret $\phi_k$ versus the round number $k$. The number of job positions is $b=5$ for $n=100$ candidates per round. Score distribution: uniform, $S_{i,k} \sim \mathcal{U}(0,1)$ (top row); exponential, $S_{i,k} \sim \text{Exp}(1), \, \forall i, \, \forall k$ (bottom row). The number of initially empty job positions is $r=0$ (left) and $r=b=5$ (right). Purple lines use \CCMDP in full-information (plain line), partial-information (dashed line) and no-information (dotted line) settings.}
\label{fig:dynamic_prog}
\end{figure}

In \Fig{fig:dynamic_prog}, we also show the performance of the algorithms for cases with different levels of \DM's knowledge for the score distribution $f_S$. In the \emph{full-information} case (plain line), $f_S$ is perfectly known. In the \emph{partial information} case (dashed line), \DM knows only the shape of the $f_S$ (\eg uniform, normal, exponential, \etc) and needs to learn its parameters (\eg lower and upper bounds, mean, \etc). Finally, the \emph{no-information} case (dotted line) is when the \DM does not hold any information about $f_S$ (\eg the shape, as discussed in the example of \Sec{sec:unif}), or not even a way to compute an absolute score per candidate. In that case, the \DM should rather rely on relative ranks that are re-assessed after the examination of each individual (see \Sec{sec:unif}).

We observe that, provided the \DM knows the shape of the score distribution (\ie partial-information), the learning of its parameters throughout the rounds is relatively fast, and the \regret quickly converges to that of full-information (plain lines). Here, the strategy is much slower in converging towards the full-information case, which is achieved for $r=0$ after approximately 40 rounds. The slow convergence is due to the \DM's inability to estimate each candidate's absolute rank before having examined all other candidates (see \Eq{eq:rel_rank}).

\section{Conclusion and discussion}
In this paper we presented a new algorithm, called \emph{\CCMDPfull} (\CCMDP), for the \WSSPfull (\WSSP), considering the case where the incoming candidates have scores following a known distribution. The proposed algorithm is based on a dynamic programming approach and achieves optimal threshold estimation at each step of the sequence of interviewed candidates. Experiments have been performed in the multi-round setting, which is interesting for real-world reoccurring recruitment processes. \CCMDP demonstrated a clearly better performance than existing algorithms, regardless the number of initially empty job positions. We additionally proposed a rank-based dynamic programming alternative that can go beyond the need of knowing perfectly the distribution that generates the scores, yet, resulting in satisfying outcomes. 

\bibliographystyle{IEEEtranS} 
\bibliography{ICTAI19_arxiv}

\end{document}


\onecolumn{
\title{\textbf{Appendix}\\
{\Large {Optimal Multiple Stopping Rule for Warm-starting Sequential Selection}}}

\author{Mathilde Fekom \quad Nicolas Vayatis \quad  Argyris Kalogeratos
\\
{\small CMLA -- ENS Paris-Saclay, 94230 Cachan, France}
\\ 
{\small\texttt{\{fekom,\,vayatis,\,kalogeratos\}@cmla.ens-cachan.fr}}
}
\maketitle
\noindent Recall that for a for an \iid score-based distribution, when the objective function is to maximize the sum of the selected scores,  the following relation holds:
\begin{align}
V^{\Xj,\Yj}_{j} &= \Exp{\max\big(V^{\Xj,\Yj}_{j+1}, S_{j}+ \max(V^{\Xj-1,\Yj}_{j+1}, V^{\Xj,\Yj-1}_{j+1})\big)}, \, \forall j \in \{1,...n\}.
\label{eq:proof_recurrence}
\end{align}
\inlinetitle{For a fixed distribution $S_j \in [\alpha,\beta], \, \forall j \leq n$, and $\mu = \Exp{S_j}$}{.}
\begin{proposition} Let a \WSSP with a population of \iid scores $S \in [\alpha, \beta]$, each following a given distribution $f_S$. Set $F_S(x) = \int^x_\alpha f_S(x)$ to be its cumulative distribution function.
The situation is as follows, $j-1$ interviews have been processed, $\Xj \in \{0,...,r\}$ job positions are empty, and $\Yj \in \{0,...,b-r\}$ are occupied by preselected employees.
The recurrence relation in \Eq{eq:proof_recurrence} becomes: 
\begin{equation}
V^{\Xj,\Yj}_{j} - V^{\Xj,\Yj}_{j+1} = Z_{j+1}^{\Xj,\Yj}\big(F_S(Z_{j+1}^{\Xj,\Yj})-1\big)+ \int_{Z_{j+1}^{\Xj,\Yj}}^{\beta} s f_S(s) ds,
\end{equation}
where $Z_{j+1}^{\Xj,\Yj} := V^{\Xj,\Yj}_{j+1} - \max(V^{\Xj-1,\Yj}_{j+1}, V^{\Xj,\Yj-1}_{j+1})$.
\label{prop:proof_relation}
\end{proposition}
\vspace{4mm}
\begin{proof}
Let each candidate have an \iid score $S_j\in \real, \, \forall j\leq n$. The objective function to maximize is the sum of the scores of the individuals with a job position at the end of the selection.

We have:
\begin{align}
V^{\Xj,\Yj}_{j} &= \Exp{\max\big(V^{\Xj,\Yj}_{j+1}, S_{j}+ \max(V^{\Xj-1,\Yj}_{j+1}, V^{\Xj,\Yj-1}_{j+1})\big)} \\
V^{\Xj,\Yj}_{j} &= V^{\Xj,\Yj}_{j+1} \Prob \big(V^{\Xj,\Yj}_{j+1} - \max(V^{\Xj-1,\Yj}_{j+1}, V^{\Xj,\Yj-1}_{j+1}) \geq S_j \big) + \max(V^{\Xj-1,\Yj}_{j+1}, V^{\Xj,\Yj-1}_{j+1})\Prob \big(V^{\Xj,\Yj}_{j+1} - \max(V^{\Xj-1,\Yj}_{j+1}, V^{\Xj,\Yj-1}_{j+1}) < S_j \big)\\
& + \int_{V^{\Xj,\Yj}_{j+1} - \max(V^{\Xj-1,\Yj}_{j+1}, V^{\Xj,\Yj-1}_{j+1})}^\beta s f_S(s) ds\\
\intertext{Set $Z_{j+1}^{\Xj,\Yj} := V^{\Xj,\Yj}_{j+1} - \max(V^{\Xj-1,\Yj}_{j+1}, V^{\Xj,\Yj-1}_{j+1})$, and $F_S(x) = \int^x_\alpha f_S(x)$ the cumulative density function:}
V^{\Xj,\Yj}_{j} &=   V^{\Xj,\Yj}_{j+1}F_S(Z^{\Xj,\Yj}_{j+1}) + \max(V^{\Xj-1,\Yj}_{j+1}, V^{\Xj,\Yj-1}_{j+1})(1-F_S(Z^{\Xj,\Yj}_{j+1})) + \int_{Z^{\Xj,\Yj}_{j+1}}^\beta s f_S(s) ds\\
V^{\Xj,\Yj}_{j}  &= V^{\Xj,\Yj}_{j+1} - Z^{\Xj,\Yj}_{j+1} +Z^{\Xj,\Yj}_{j+1}F_S(Z^{\Xj,\Yj}_{j+1})+ \int_{Z^{\Xj,\Yj}_{j+1}}^\beta s f_S(s) ds\\
V^{\Xj,\Yj}_{j}  &= V^{\Xj,\Yj}_{j+1} + Z^{\Xj,\Yj}_{j+1} \big(F_S(Z^{\Xj,\Yj}_{j+1})-1\big)+ \int_{Z^{\Xj,\Yj}_{j+1}}^\beta s f_S(s) ds.
\end{align}
\end{proof}
%

\inlinetitle{Recurrence relation for a uniform distribution $S_j \sim U(\alpha,\beta), \, \forall j \leq n, \, \mu = (\alpha+\beta)/2$}{.}
%
\begin{proposition} Set $S_j \sim \mathcal{U}(\alpha,\beta), \, \forall j \leq n$. \Proposition{prop:proof_relation} becomes:
\begin{equation}
V^{\Xj,\Yj}_{j} - V^{\Xj,\Yj}_{j+1} = \frac{{Z^{\Xj,\Yj}_{j+1}}^2- 2\alpha Z^{\Xj,\Yj}_{j+1}+\beta^2 }{2(\beta-\alpha)} - Z^{\Xj,\Yj}_{j+1},
\end{equation}
where $Z_{j+1}^{\Xj,\Yj}= V^{\Xj,\Yj}_{j+1} - \max(V^{\Xj-1,\Yj}_{j+1}, V^{\Xj,\Yj-1}_{j+1})$.
\label{prop:proof_unif}
\end{proposition}
\begin{proof} We have $F_S(s)=\frac{s-\alpha}{\beta-\alpha}$ and $f_S(s) = \frac{1}{\beta-\alpha}$:
\begin{align}
V^{\Xj,\Yj}_{j}  &= V^{\Xj,\Yj}_{j+1} + Z^{\Xj,\Yj}_{j+1} \left(\frac{Z^{\Xj,\Yj}_{j+1}-\alpha}{\beta-\alpha} -1\right)+ \frac{1}{\beta-\alpha}\int_{Z^{\Xj,\Yj}_{j+1}}^\beta s ds\\
V^{\Xj,\Yj}_{j} &= V^{\Xj,\Yj}_{j+1} - Z^{\Xj,\Yj}_{j+1} + \left(\frac{{Z^{\Xj,\Yj}_{j+1}}^2-\alpha Z^{\Xj,\Yj}_{j+1}}{\beta-\alpha}\right) +\frac{\beta^2 - {Z^{\Xj,\Yj}_{j+1}}^2}{2(\beta-\alpha)} \\
V^{\Xj,\Yj}_{j} &= V^{\Xj,\Yj}_{j+1} - Z^{\Xj,\Yj}_{j+1} + \frac{{Z^{\Xj,\Yj}_{j+1}}^2}{2(\beta-\alpha)} +\frac{\beta^2 - 2\alpha Z^{\Xj,\Yj}_{j+1}}{2(\beta-\alpha)} \\
V^{\Xj,\Yj}_{j} &= V^{\Xj,\Yj}_{j+1} - Z^{\Xj,\Yj}_{j+1} +\frac{{Z^{\Xj,\Yj}_{j+1}}^2- 2\alpha Z^{\Xj,\Yj}_{j+1}+\beta^2 }{2(\beta-\alpha)}.
\end{align}
\end{proof}
\inlinetitle{Recurrence relation for a discrete uniform distribution $S_j \sim U\{\alpha,\beta\}, \, \forall j \leq n, \, \mu = (\alpha+\beta)/2$}{.}
%
The objective is to minimize the sum of the scores, that can be seen as ranks.
\begin{align}
V^{\nres, b-\nres}_{n-\nres+1} &= \Exp{\sum_{j=n-\nres+1}^{n} S_j + \sum_{i=1}^{b-\nres}\So{i} } = \nres \mu +  \sum_{i=1}^{b-\nres}\Exp{\So{i}}\\
\intertext{Note that here, $Z^{\Xj,\Yj}_{j+1} = V^{\Xj,\Yj}_{j+1} - \min(V^{\Xj-1,\Yj}_{j+1}, V^{\Xj,\Yj-1}_{j+1})$. Hence:}
V^{\Xj,\Yj}_{j}  &= V^{\Xj,\Yj}_{j+1}(1-F_S(Z^{\Xj,\Yj}_{j+1})) +( V^{\Xj,\Yj}_{j+1}- Z^{\Xj,\Yj}_{j+1})F_S(Z^{\Xj,\Yj}_{j+1}) + f_S\frac{Z^{\Xj,\Yj}_{j+1}(Z^{\Xj,\Yj}_{j+1}+1)}{2}\\
V^{\Xj,\Yj}_{j}  &= V^{\Xj,\Yj}_{j+1} - Z^{\Xj,\Yj}_{j+1}F_S(Z^{\Xj,\Yj}_{j+1}) + f_S\frac{ {Z^{\Xj,\Yj}_{j+1}}^2+Z^{\Xj,\Yj}_{j+1} }{2}.\\
\intertext{More specifically, for ranks between $\alpha=1$ and $\beta=n+b$, we have $F_S(k) = \frac{k}{n+b}$, and $f=\frac{1}{n+b}$, thus:}
V^{\Xj,\Yj}_{j}  &= V^{\Xj,\Yj}_{j+1} - \frac{ {Z^{\Xj,\Yj}_{j+1}}^2 }{n+b} + \frac{1}{n+b}\frac{{Z^{\Xj,\Yj}_{j+1}}^2+Z^{\Xj,\Yj}_{j+1}}{2}\\
V^{\Xj,\Yj}_{j}  &= V^{\Xj,\Yj}_{j+1} - \frac{ {Z^{\Xj,\Yj}_{j+1}}^2 -Z^{\Xj,\Yj}_{j+1} }{2(n+b)}.
\end{align}
\\

\inlinetitle{Recurrence relation for an exponential distribution $S_j \sim \text{Exp}(\lambda), \, \forall j \leq n, \, \mu = 1/\lambda$}{.}

We have, $\alpha =0$, $\beta =+\infty$, $F_S(s) = 1-e^{-\lambda s}$ and the truncated expectation is \st:
\begin{align}
\int_{Z^{\Xj,\Yj}_{j+1}}^\beta s f_S(s) ds &= \int_{Z^{\Xj,\Yj}_{j+1}}^{+\infty} s \lambda e^{-\lambda s}\\
\int_{Z^{\Xj,\Yj}_{j+1}}^\beta s f_S(s) ds&=Z^{\Xj,\Yj}_{j+1} e^{-\lambda Z^{\Xj,\Yj}_{j+1}}  + \frac{e^{-\lambda Z^{\Xj,\Yj}_{j+1}}}{\lambda}.
\intertext{Hence:}
V^{\Xj,\Yj}_{j}  &= V^{\Xj,\Yj}_{j+1} - Z^{\Xj,\Yj}_{j+1}e^{-\lambda Z^{\Xj,\Yj}_{j+1} }+Z^{\Xj,\Yj}_{j+1} e^{-\lambda Z^{\Xj,\Yj}_{j+1}}  + \frac{e^{-\lambda Z^{\Xj,\Yj}_{j+1}}}{\lambda}\\
V^{\Xj,\Yj}_{j}  &= V^{\Xj,\Yj}_{j+1} + \frac{e^{-\lambda Z^{\Xj,\Yj}_{j+1}}}{\lambda}.
\end{align}